\documentclass[conference]{IEEEtran}
\usepackage[numbers]{natbib}
\usepackage{hyperref}
\usepackage{tikz}
\usepackage{graphicx}

\hypersetup{
  pdftitle={Data-Intensive Supercomputing in the Cloud},
  pdfauthor={Warren et al.},
  colorlinks=false
}

\begin{document}

\title{Data-Intensive Supercomputing in the Cloud: \\ Global Analytics for Satellite Imagery \\[-.1in]
{\small \em In Proceedings of the 7th International Workshop on Data-Intensive Computing in the Cloud (DataCloud '16). IEEE Press, 24-31, 2016}\\[-.15in]}

\author{\IEEEauthorblockN{Michael S. Warren,
    Samuel W. Skillman, \\
    Rick Chartrand,
    Tim Kelton, Ryan Keisler, David Raleigh \\[.3em]}
  \IEEEauthorblockA{Descartes Labs \\
    1925 Trinity Drive \\
    Los Alamos, NM 87544}
\and
\IEEEauthorblockN{Matthew Turk \\[.3em]}
\IEEEauthorblockA{School of Information Sciences\\
  University of Illinois Urbana-Champaign\\
  Urbana, IL 61801}
}

\maketitle

\begin{abstract}

  We present our experiences using cloud computing to support
  data-intensive analytics on satellite imagery for commercial
  applications.  Drawing from our background in high-performance
  computing, we draw parallels between the early days of clustered
  computing systems and the current state of cloud computing and its
  potential to disrupt the HPC market.  Using our own virtual file
  system layer on top of cloud remote object storage, we demonstrate
  aggregate read bandwidth of 230 gigabytes per second using 512
  Google Compute Engine (GCE) nodes accessing a USA multi-region standard storage bucket.  This
  figure is comparable to the best HPC storage systems in existence.
  We also present several of our application results, including the
  identification of field boundaries in Ukraine, and
  the generation of a global cloud-free base layer from Landsat imagery.
\end{abstract}

\vspace{1em}
\noindent

\section{Introduction}

One of the most rapid and remarkable technological advances in history
occurred with the steam engine after Watt's patent expired in 1800.
Over a 30 year period, engineers improved the efficiency of the
Cornish high-pressure steam engine by a factor of five~\cite{FM1284} (a
compounded improvement of 5.5\% per year).  Among computing and
information systems today, such a rate of change would be considered
very slow.  Against the present background of Moore's
Law~\cite{moore_1965} (loosely representing a doubling in performance
every 18 months, or 60\% improvement per year) the technological rate
of change in computing is 10 times higher than the peak of the
industrial revolution.  As software replaces physics as the system of
the world, these rates of exponential change are becoming apparent in ever-larger
aspects of people's day-to-day experience.

\begin{figure}[htb]
  \centering
  \resizebox{\columnwidth}{!}{\includegraphics{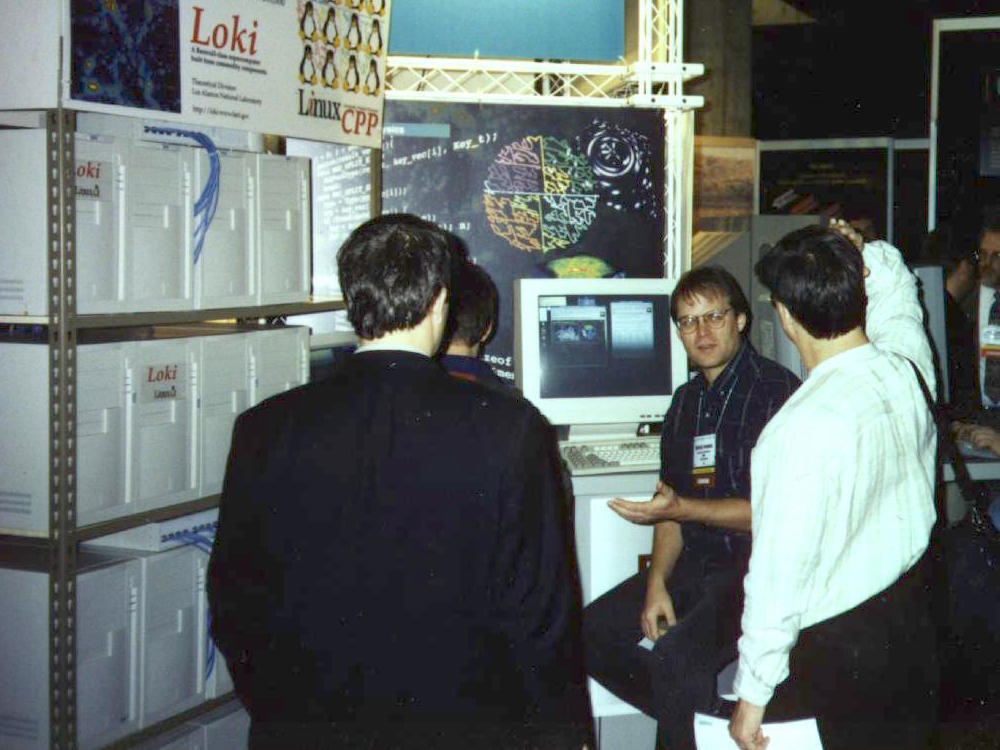}}
  \caption{This photo shows Mike Warren extolling the virtues of
    commodity parallel computing on the exhibition floor of
    Supercomputing '96 in Pittsburgh.  On the left, the 16 processor
    Loki Beowulf cluster was the first demonstration of a Linux
    cluster at the annual Supercomputing conference. This cluster won
    the Gordon Bell price/performance prize the next
    year~\cite{warren_pentium_1997}.  The descendants of this Intel/Linux
    architecture have come to dominate both high-performance computing
    and the cloud.}
  \label{fig:sc96}
\end{figure}

Twenty years ago, we were the first to demonstrate a commodity
computing cluster (Figure~\ref{fig:sc96}) that was
competitive with special-purpose HPC architectures~\cite{warren97b,warren_pentium_1997} and by some
accounts~\cite{bell_whats_2002}, led to the demise of the traditional
supercomputer.  The effort to build our own cluster was a minor
perturbation on the effort that had already been expended writing the
application software.  Today, the cloud is poised to play the same
disruptive role with decreasing costs, vanishing barriers to entry,
rapidly increasing performance, a stable programming interface, and a
rich ecosystem of open-source software libraries and applications to
build upon.

\section{Origins of the Cloud}

Loki and Hyglac~\cite{warren_pentium_1997} were clusters of Intel processors
with Ethernet as a communication fabric, using the Linux operating
system, constructed in 1996.  The same triad of Intel/Linux/Ethernet describes the majority
of cloud computing systems today, and if we allow more exotic
communication networks, it describes most of the fastest
supercomputers in the world.  While other groups were investigating the
potential of commodity clusters in the same era~\cite{anderson1995case}, our commercial off-the-shelf (COTS)
approach distinguished itself through winning the Gordon Bell price/performance prize in 1997.
It is also notable that we built the
first machine on the TOP500 list which used Linux as the OS in 1998
\url{www.top500.org/system/166764}.  Today, 99\% of the TOP500
supercomputers run Linux
\url{www.top500.org/statistics/details/osfam/1}, not to mention its
prevalence in other things, from electric cars to nano-satellites~\cite{boshuizen_results_2014}.  This
domination is not accidental; the hardware, operating system and
network APIs established at all levels by the Intel/Linux pairing
(with the world wide web at WAN scales) that came together in the
mid-90s has provided a stable base computing environment
for two decades.

Following up the initial demonstration of the BEOWULF
project~\cite{becker_beowulf:_1995}, a clear outcome of this work was
the democratization of access to HPC resources.  A technical group
could buy and assemble the components to make their own cluster of
machines tailored to their own problem domain.  Large data-analysis problems
were also amenable to this approach~\cite{warren2007astronomical}. However, as the price
per CPU continued to fall, the scale of the human effort required to
assemble, house, power and maintain such a machine became more than a
minor effort.  While it was still possible to reach performance among
the top 100 supercomputers with mail-order parts and part-time system
management in 2002~\cite{warren_space_2003}, today it requires a machine
with over 30,000 cores.  Thus, economies of scale become
a dominant factor for significant computing needs, and the cloud
has naturally evolved from the fertile environment created from the early
work with commodity clusters.  The promise of \cite{warren98a} ``The
software community that develops around these machines (if it remains
open and follows the Linux development philosophy) will allow users,
vendors, and the research community to each contribute to a robust
software environment'' has been realized.

There are certainly scientific computing problems outside
the realm of cloud computing as it exists today, when dedicated
low-latency communication and extreme scaling of tightly-coupled
processes are required.  Our past work in computational cosmology is
one exemplar~\cite{skillman_dark_2014}.  Even so, it is likely that a
code which can scale efficiently to more than $10^5$ processors on a
dedicated supercomputer will perform adequately for smaller problems
on a few thousand processors in the cloud.  There is also no clear
barrier (other than economic) preventing the deployment of HPC technology
such as low-latency high-performance networks into the cloud in the
future.

While being a ``utility" is one of the greatest features of cloud
computing, it also is one of the factors that could slow its
acceptance.  Big computers are institutional status symbols.  They
serve as an impressive physical artifact to show to visitors who might
otherwise have difficulty appreciating the intangible properties of
a revolutionary algorithm or well-written software.
Large computers grow whole ecosystems around themselves, involving
procurement, machine rooms, electrical power and support staff.  All
of those local sources of influence are disrupted in a move to cloud
computing, where all that remains is an optical fiber and a monthly
bill. 

\section{Architectural Considerations}

A major advantage of the cloud is almost instant access to vast
amounts of computing.  In our experience it is the rule rather than
the exception that thousands of cores can be spun up and running code
within a minute or two.  This is in stark contrast to a typical shared
supercomputer where jobs can often sit in the queue for hours or days.
When time-to-solution matters, it matters when a job is done, not how
fast it runs after it starts.

\begin{figure*}[bt]
  \centering
  \includegraphics[width=13cm]{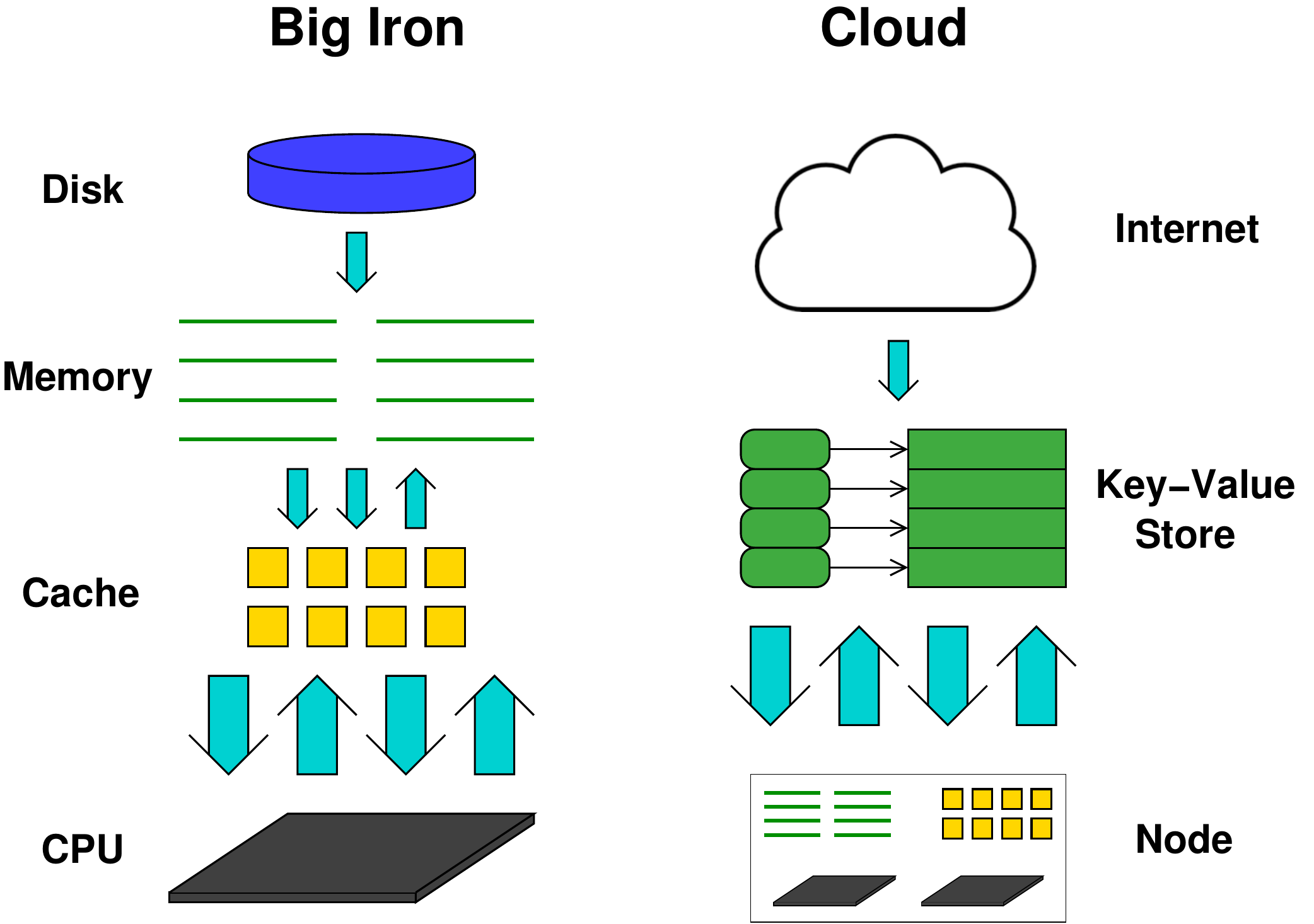}
  \caption{Many of the hard-learned lessons of the past decades of
    supercomputing are still relevant to cloud computing, with a
    simple translation between architectural components (memory becomes
    a key-value store, and a CPU becomes the entire compute node).  Where
    efficiency on a ``Big Iron'' supercomputer is most often dependent
    on dealing effectively with the limited bandwidth available from
    main memory, the equivalent limitation in the cloud is the network
    connection between the processing nodes and data storage.  The
    same techniques developed for ``cache-friendly'' algorithms to increase
    data locality can often be adapted for use in cloud architectures.
  }
  \label{fig:cloudcompute}
\end{figure*}

From an economic point of view, Table~\ref{table:cost} demonstrates the
fundamental lesson that the cost of a programmer must be amortized
over an enormous amount of computing.  As the price of computing
continues to fall, every programmer must contribute to a system which
scales to larger and larger processor counts.  The yearly salary of an
average programmer is now equivalent to 1000 pre-emptible cores
running 24 hours per day, 365 days per year.  In that context, for any
business not utilizing thousands of cores constantly, computing is
already free.  The converse is that programmers must be as productive
as possible.  Existing code or libraries that are suited to the task at
hand must be used as often as possible, and anything that breaks
existing code must be perceived as an enormous expense.  Further,
understanding the architectural constraints and best approaches for optimization
in overall system design requires an additional level of insight and experience
(Figure~\ref{fig:cloudcompute}).

\begin{table}
\hfil
\begin{tabular}{|l|l|r|}
\hline  2016 Cost (\$/s) & Unit & Description\\ \hline
\hline
$1.0 \cdot 10^{-8}$ &  Gigabyte & Cloud storage \\
$1.5 \cdot 10^{-8}$ &  Gigabyte & Persistent magnetic disk \\
$6.5 \cdot 10^{-8}$ &  Gigabyte & Node solid state disk \\
$1.6 \cdot 10^{-7}$ &  Gigaflop/s & LINPACK 64-bit floating point \\
$2.5 \cdot 10^{-7}$ &  Gigabyte & Node memory \\
$3.8 \cdot 10^{-5}$ &  Gigabyte/s & Local network \\
$1.0 \cdot 10^{-2}$ &  Gigabyte/s & to Wide Area Network \\
$2.8 \cdot 10^{-2}$ &  & Skilled human labor\\
$1.0 \cdot 10^{-1}$ &  Gigabyte/s & to Public Internet \\
\hline
\end{tabular}\\
\caption{Fundamental computing costs.  Figures are derived from
  published Google Cloud Platform pre-emptible node
  pricing and our own performance
  measurements in September, 2016.  Bandwidths and capacities have
  been converted to a cost
  per second per giga-unit to facilitate total cost estimation.  For
  example, storing one Petabyte (1 million gigabytes) for one year
  (31.5 million seconds) in Cloud Storage costs \$315,000.  One dollar
  can currently buy ~60 seconds of programming labor, deliver 10
  Gigabytes to the Internet, store 46 gigabytes in DRAM for 1 day, or provide
  $6 \times 10^{15}$ floating point operations.
}
\label{table:cost}
\end{table}

\subsection{Everything is a File}

While the cloud solves many of the practical difficulties around
having access to computing capability, it does not make architecting
and developing application software an easier problem.  A particularly
large stumbling block for large data-processing problems is the object
storage API.  One of the defining features of Unix and Unix-like
operating systems is that ``everything is a file."  Object storage
does not conform to this fundamental interface, and thereby breaks a
vast number of tools, utilities, libraries and application code.  For
much the same reason, we do not consider the
Hadoop~\cite{white2012hadoop} ecosystem to be the right model for our
data-intensive computing tasks, although this is mostly the fault of
HDFS.  It is possible that alternate approaches such as Apache Spark
could contribute to our software infrastructure, although we have not
yet deployed any solutions using that framework.

Persistent storage in the cloud is generally made available through a
RESTful interface where the traditional POSIX open/read/close are
instead performed with a HTTP \verb-GET- operation, while
open/write/close is a \verb-PUT- operation.  The hierarchy of a
traditional file system is replaced by globally unique identifier.
Updating the data in an object requires it to be re-written in its
entirety.  These interface constraints allow object storage to be
more scalable and able to be implemented less expensively than a
traditional file system.  An additional enormous benefit is that it
largely eliminates the complexities associated with data locality,
since all data is remote.  The price to be paid is higher access
latency, but much worse is the additional programming effort to modify
code to work with an object store.  Often it is easier to copy the
data from an object store to an intermediate file on the local
filesystem, process it using the normal POSIX I/O interface, and then
copy the output back to object storage.

For moderate file sizes and low I/O rates, making a copy in local
storage can perform reasonably well.  The cost of the second read
(from the intermediate file into application memory) is usually mostly
hidden since the file is cached in memory by the operating system.
For larger files at higher data rates, this breaks down and we will
run into bandwidth restrictions at the local storage level.  Using GCE
as an example, standard local storage has a limit of 180 MB/s of read
bandwidth and 120 MB/s of write bandwidth.  We are then in the
paradoxical position of being able to read from a remote object store
at a higher rate than from the (virtual) local disk.

As an alternative method, it is possible to read from the object store
directly into memory.  This can work, but it is also subject to
difficulties.  If the application reads directly into userspace
memory, the ability of the operating system to share those data blocks
with other processes is lost. (This sharing happens naturally via the
filesystem interface).  Additionally, many common programming library
interfaces have not been designed to work with a memory interface (for
instance, they expect a file name or file pointer, with the actual
memory access being private to the library).

When objects are large (approaching the size of the memory available
to each processor) and existing code has been designed to read smaller
portions of a file and process them individually, there are no good
alternatives for interacting with object storage.  The data must be
accessed via an interface that provides for random read access.  The
currently favored method for mapping the file abstraction is to use a
virtual file system, such as FUSE.

\subsection{Festivus}

FUSE (Filesystem in Userspace) is an interface for userspace programs
to export a filesystem to the Linux kernel.  For the case of object
storage, the job of FUSE is to translate things the kernel understands
--- system calls like open and read and identifiers like inodes or
file names --- to things the object store understands --- actions like
HTTP GET and identifiers like a Uniform Resource Locator (URL).
Several implementations of FUSE for cloud object storage have
been implemented and are in use (s3fs, gcsfuse).

None of the existing FUSE implementations met our needs.  The most
significant problems were sub-optimal read performance and slow
metadata access.  Given the importance of a high-performance file system
interface on top of cloud object storage, we felt developing our own
low-level asynchronous FUSE interface was worth the effort.

Our FUSE implementation is named festivus (a file system for the rest
of us).  It has been written from scratch using the low-level threaded asynchronous API of libfuse.
libfuse provides the implementation for userspace communication with the FUSE kernel module,
reading requests from the kernel which are passed to festivus using callbacks, and
then returning the festivus responses to the kernel.
Rather than query the object store itself for object metadata, we
maintain our own separate scalable in-memory key/value store to
perform metadata-related operations (this metadata server is shared by
all instances of the file system).  We currently use
Redis~\cite{carlson_redis_2013}, although similar functionality could
be provided by other systems.  An important optimization for object
data access is to increase the Linux kernel parameter
\verb-FUSE_MAX_PAGES_PER_REQ- from its default value of 32 (which
limits read chunks to 128k) to a much higher value (the results
presented here increase this value to 1024 pages; 4 MB on the Intel
Xeon).  With this change, the \verb-VM_MAX_READAHEAD- kernel parameter
must also be modified appropriately.  This optimization requires the
modified kernel to be installed on any node running the festivus
filesystem.

In addition to large distributed processing tasks, many pieces of
production software leverage the POSIX file system.  One such
application is Mapserver, an open source software package for
publishing spatial data and interactive mapping over http~\cite{vatsavai2006umn}.
Traditionally, Mapserver is configured to serve data through its local
file system, limiting the amount of imagery per node that can be
served.  Festivus allows processed, compressed, and tiled imagery
stored in Google Cloud Storage to be served through the POSIX interface
using Mapserver the same as if the data was served natively on
attached disks.  This allows hundreds of terabytes of imagery to be
served on a single instance.  Traditional mechanisms for scaling
Mapserver would involve replicating the datasets for a
specific set of imagery or leveraging large dedicated NFS systems.
Mapserver on top of Festivus allows for horizontal scaling without
further replication of data while leveraging the economies of scale
and ease of data management provided by cloud object storage.

\subsection{Domain Decomposition}

When computing in parallel, a task that must be done well in order to
succeed is splitting up the data among processors.  This is often
referred to as {\em domain decomposition}.  A single image of the
Earth with pixel scales less than about 10km is too large to process
efficiently, so the image must be ``tiled", or split into pieces that
can be processed independently.  For current computer architectures
and memory storage capacities, a reasonable size for image tiles would
be between 256 x 256 and 4096 x 4096 pixels, depending on the
application.  Note that these tiles are not necessarily individual
files, since some image formats support internal tiling, or even
further, the API layer may provide virtual tiles which are constructed
on-the-fly from the underlying data.

Two common map projections that represent the spherical surface of
the Earth as a regular grid are the UTM (Universal Transverse
Mercator) projection, and the Web Mercator projection.  The Web
Mercator projection is easily tiled, because the image dimensions are
precisely a power of two in both co-ordinates.  The level of the
decomposition ($L$) divides the world into $4^L$ pieces.  An appropriate
level can be chosen to satisfy various constraints.  For instance, that a
number of time slices for a given tile can fit into processor
memory at one time.  Web Mercator is ubiquitous for simple map
interfaces, but can not be used for anything beyond simple analysis
because the pixel areas are not equal.  As a pixel becomes farther
from the equator, it represents a smaller and smaller area on the
surface of the Earth.  Web Mercator has been declared unacceptable for
official use by the US government
\url{http://earth-info.nga.mil/GandG/wgs84/web_mercator/index.html}

The UTM projection is not as simple.  UTM first splits the world into
60 zones, and within each zone pixels are split into nearly equal
areas referenced by their ``x" or ``Easting" co-ordinate and their ``y"
or ``Northing" co-ordinate.  All UTM distances are measured in
meters. The number of pixels which span a zone in the East-West
direction depends on the distance from the equator.

For the most efficient and accurate processing of multiple datasets,
they should share a common co-ordinate reference system.  Since
operations to interpolate pixels to a different map projection or
resolution can affect the data quality and require additional
computational resources, we seek to minimize the number of such
operations.  This suggests using UTM as the common map projection,
since most data is delivered in UTM co-ordinates.

The UTM tiling system we have created is defined by a number of
parameters.  It is applied to each of the 60 UTM zones with identical
parameters, with the zone designated by $z$.  A similar construction
can be applied to the polar UPS projection. The parameters are the
origin of the tiling system, the number of pixels in the tile $x$ and
$y$ dimension, the border (overlap), and the spatial resolution of the
pixels.  Since a UTM zone is 6 degrees across, that represents 668 km
at the equator.  For pixel scales larger than about 200 meters, a
single tile will cover the east-west extent of a UTM zone.  For
smaller pixel scales, multiple tiles are required.  For 10m
resolution, such as the Sentinel-2 satellite, 17 4096-pixel wide tiles
would be required.

In the y-dimension, the distance from the equator to the pole is near
10000km, so the number of 4096 x 4096 tiles to span that distance is
about 10 for a 250m pixel tile, or 244 for a 10m tile.  The southern
hemisphere can be handled with a similar number of tiles using a
negative index referenced from the equator, or referenced by their
northing co-ordinate from the south pole using the southern ``S"
designator for the zone.

We store our pre-processed imagery using the JPEG 2000
standard~\cite{jtc1sc29-2004-jpeg2000, taubman-2001-jpeg2000} due to
its significant advantages in terms of compression and image types as
well as its support for internal tiling and a scalable
multi-resolution codestream that can be ordered to best fit
applications demands.  We also use a small portion of the ``Part 2''
extensions defining the more flexible JPX file format.

\section{Performance}

\begin{figure}[htb]
  \centering
  \resizebox{\columnwidth}{!}{\includegraphics{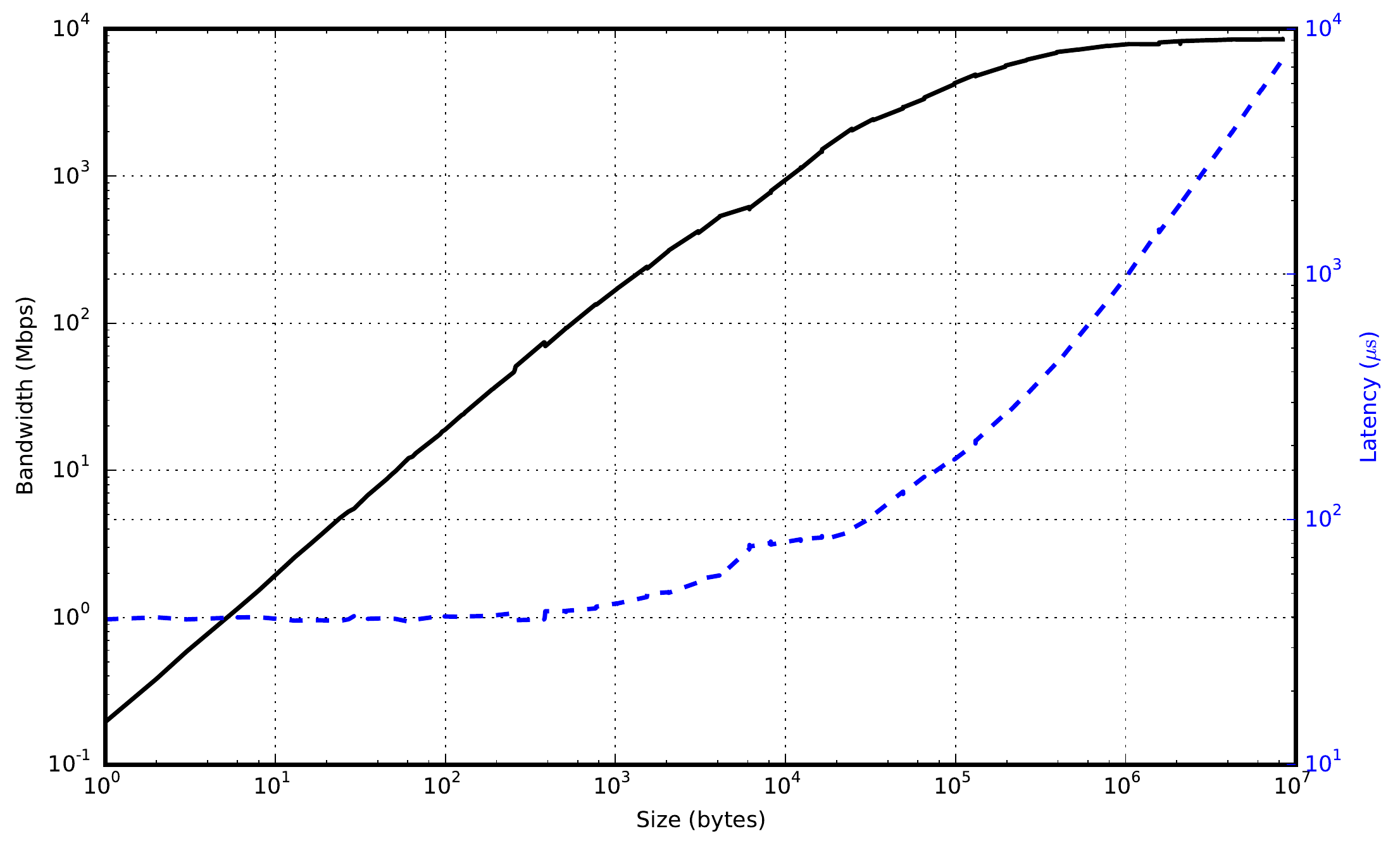}}
  \caption{Bandwidth (solid) and latency (dashed) for a single thread on a 16-vCPU Google Compute Engine node, plotted against message size
    on the x-axis.  Small message latency
    is about 40 microseconds, while large message bandwidth reaches 8.6 Gigabits/second.  For multiple
    threads total bandwidth reaches 16 Gigabits/second.
    Compared to similar measurements from 20 years ago~\cite{warren97b} message latency has improved a factor of 5
    and bandwidth per HW core by about a factor of 20.}
  \label{fig:netpipe}
\end{figure}

\begin{table}[htb]
\begin{center}
\begin{tabular}{|l|r|r|r|}
\hline
& Cluster (2003) & Cray XE6 & Cloud \\
\cline{2-4}
& \multicolumn{3}{|c|}{Performance per core} \\
\hline
STREAM copy  (MB/s)  & 1203.5 & 2612.0 & 1753.9 \\
STREAM add    & 1237.2 & 2591.7 & 1731.4 \\
STREAM scale  & 1201.8 & 2820.3 & 1923.3 \\
STREAM triad  & 1238.2 & 2857.9 & 1953.2 \\
\hline
NPB BT  (Mop/s)    & 321.2 & 1445.1 & 1244.0\\
NPB SP      & 216.5 &  706.0 &  613.5\\
NPB LU      & 404.3 & 1599.0 & 1897.9\\
NPB MG      & 385.1 & 1134.0 &  940.9\\
NPB CG      & 313.1 &  380.2 &  299.5\\
NPB FT      & 351.0 &  767.5 &  923.6\\
\hline
LINPACK (Gflop/s) & 3.30 & 14.94 & 25.46 \\
\hline
\end{tabular}\\[1em]




\caption{A comparison of 2003 era Beowulf cluster Gordon Bell
  price/performance prize finalist~\cite{warren_space_2003}
  to a Cray XE6 node and 2015 Cloud performance.  The top
  set of results is memory bandwidth measured in Mbytes/s by the
  STREAM benchmark~\cite{mccalpin_memory_1995}, middle is from the
  Class C NAS Parallel Benchmarks~\cite{bailey_nas_1991} in Mop/s, and
  last line is LINPACK Gflop/s. Beowulf results are per CPU on a 2.53
  GHz Intel Pentium 4. Cloud results are per HW core on a 16-core 2.3
  GHz Intel Haswell node. Cray results are per Bulldozer compute unit
  of a dual-socket XE6 node with 2 AMD Interlagos model 6276 2.3 GHz
  CPUs (16 cores total).  These data make it clear that performance
  per core has progressed little in the past 12 years for un-tuned
  applications (between factors of 1 and 8, with most of these
  benchmarks around a factor of 3).}

\label{tab:bench}
\end{center}
\end{table}

\subsection{1.21 Teraflops for \$1/hr} 

ASCI Red, the first Teraflop supercomputer, was deployed at Sandia National
Laboratories for a cost of \$46M (1996 dollars). Its processors were upgraded
in 1999. Disregarding operational costs (which can be substantial), that
corresponded to \$1,749 per teraflop-hour. We have recently run the LINPACK
benchmark on two pre-emptible 64-vCPU GCE nodes (\verb~n1-highcpu-64~, 32 hardware
cores per node and 57.6 GB of memory), and obtained a result of
$1.21 \mathrm{teraflops}$. 
Each of the \verb~n1-highcpu-64~ nodes cost $\$0.51$ per hour, coming to a total of
just \$0.84 per teraflop hour, an improvement in price-to-performance of over
$2000 \mathrm{x}$. We should also note the incredible ease in which our virtual
teraflop cluster was both constructed and decommisioned within a span of
several hours, most of which was spent performing a scan of the LINPACK parameter
space for peak performance.

\begin{table}[htb]
\begin{center}
\begin{tabular}{|l|r|r|}
\hline
Node Type & Nodes & Bandwidth \\
 & & (GBytes/s) \\
\hline
1-vCPU  & 1   &  0.43\\
4-vCPU  & 1   &  0.85\\
16-vCPU & 1   &   1.0\\
32-vCPU & 1   &  1.44\\
16-vCPU & 4   &   4.1\\
16-vCPU & 16  &  17.4\\
16-vCPU & 64  &  36.3\\
16-vCPU & 128 &  70.5\\
16-vCPU & 512 & 231.3\\
\hline
\end{tabular}\\[1em]

\caption{Aggregate festivus bandwidth, measured in GB/s.}
\label{table:bandwidth}
\end{center}
\end{table}\ \\[-4em]

\subsection{Festivus bandwidth}

In order to test the scalability of our FUSE-based filesystem, we ran
a performance test of reading a random (different on each node) subset of our
processed Landsat 8 imagery from Google cloud storage down to a distributed set
of 512 16-vCPU \verb~n1-standard-16~ virtual machines spread over the us-central1-c
zone. We measured an aggregate incoming bandwidth to
compute nodes across the project of over 231 GB/s. In table
\ref{table:bandwidth} we show the bandwidth as we scale up from a single
us-central1-c node to 512 nodes. A 32-vCPU node reaches over 70\%
of its network capacity (Fig.~\ref{table:bandwidth}). 
In the transition from 16 to 64 nodes we observe a drop in bandwidth
per node from approximately 1 GB/s to 500 MB/s, perhaps due to sharing of
network bandwidth between nodes. We have run similar single-node benchmarks
of \verb~gcsfuse~, where we have observed a peak bandwidth of ~340 MB/s.
However, an important aspect of our festivus implementation is the ability to
read smaller ($\sim$ 1 MB) blocks of data from a larger single file.  When reading
random 4 MB chunks of data from multiple files, we observe a similar peak bandwidth
with festivus of 850 MB/s on a single node.  A similar experiment with
\verb~gcsfuse~ reveals a peak bandwidth of only 47 MB/s
(see Table \ref{table:single_node_bandwidth}).

\begin{table}[htb]
\begin{center}
\begin{tabular}{|l|r|r|}
\hline
Blocksize & festivus & gcsfuse \\
(bytes)   & (MB/s)   & (MB/s)  \\
\hline
32768    & 12.5   &  0.4   \\
65536    & 22.6   &  0.8   \\
131072   & 47.3   &  1.6   \\
262144   & 93.0   &  2.8   \\
524288   & 156.8  &  7.3   \\
1048576  & 271.0  &  13.7  \\
2097152  & 472.0  &  24.8  \\
4194304  & 852.3  &  46.7  \\
8388608  & 1046.4 &  109.5 \\
16777216 & 1248.0 &  200.3 \\
33554432 & 1593.3 &  339.7 \\
\hline
\end{tabular}\\[1em]

\caption{Single node random I/O bandwidth vs read block size. A single read
  is performed for each file, with a random offset into the file.  For random access of 4 MB chunks,
festivus outperforms gcsfuse by a factor of 18.}
\label{table:single_node_bandwidth}
\end{center}
\end{table}

\section{Applications}

\subsection{Initial Processing}

One of the first major achievements of our satellite imagery pipeline
was the processing of over one petabyte of Landsat and MODIS imagery
in under 16 hours on April 16, 2015.  This calculation is described in
detail in~\cite{warren2015seeing}.  Our input dataset consisted of
$915.52 \times 10^{12}$ bytes of Landsat data in 5693003 bzip
compressed GeoTIFF files located at \verb+gs://earthengine-public/+,
and $101.83 \times 10^{12}$ bytes of MODIS Level 1B (2QKM) band 1
(red) and 2 (near infrared) data in 613320 sz compressed HDF4 files
(collected from the NASA ftp site and stored in Google Cloud Storage),
for a total of $1017.35 \times 10^{12}$ bytes and 6306323 files.

For this
project, we used Google Compute Engine (GCE), which is the
Infrastructure as a Service (IaaS) component of Google's Cloud
Platform.  GCE became generally available in December 2013, and offers
virtual machines using KVM as the hypervisor.

The processing stages for each Landsat image include retrieving it
from Google Cloud Storage, uncompressing it, parsing the metadata,
identifying the bounding rectangle that contains valid data, cleaning
the edges of the image, converting the raw pixel information into
meaningful units (calibrated top of atmosphere reflectance using the
appropriate constants for each satellite and accounting for solar
distance and zenith angle), tiling each image, performing any
necessary co-ordinate transformations, compressing the data into
JPEG 2000 format, and storing the result back into Cloud Storage.

Several optimizations were necessary to reduce the cost of resources
involved in the computation (illuminated by Table~\ref{table:cost}).
In particular, we aggressively reduced memory usage to allow us to run
on the smallest memory (and least expensive per core) Google Compute
Engine nodes, which contain somewhat less than 2 GBytes of memory per
hardware core, and used no conventional disk storage on the compute
nodes at all (beyond the minimum 10 GB partition required to boot the
system), working entirely in memory or from the Linux tmpfs RAM disk.
Reducing memory usage is often difficult using Python's NumPy array objects, since
they often require intermediate copies during array operations.  We
also removed most of the intermediate writes to the local file system,
going from memory buffer to memory buffer between application
libraries.   Memory
bandwidth is an important factor in our overall performance, and the
measurements we provide in Table~\ref{tab:bench} are relevant for this application.

To manage the creation of asynchronous tasks for processing millions
of scenes across the worker nodes, an
asynchronous task queue approach (specifically the Python Celery library~\cite{celery})
was used.  Celery's API allows multiple asynchronous job queues to be
created, manage the list of tasks and their parameters, and insert
them into a pluggable backend key-value pair store (Redis~\cite{sanfilippo2010redis}).  As
worker nodes are provisioned and start, they connect to the Celery
broker to receive processing tasks in the queue.  Work tasks can define data location using 
cloud object storage's native location and API's, additionally, code 
utilizing POSIX file system can utilize Festivus for accessing data objects 
just as they would traditional POSIX based datasets.  This allows code 
with dependencies on POSIX based filesystems access to cloud based datasets 
and leverages the economic and scale of cloud object storage.

\begin{figure}[htb]
  \centering
  \resizebox{\columnwidth}{!}{\includegraphics{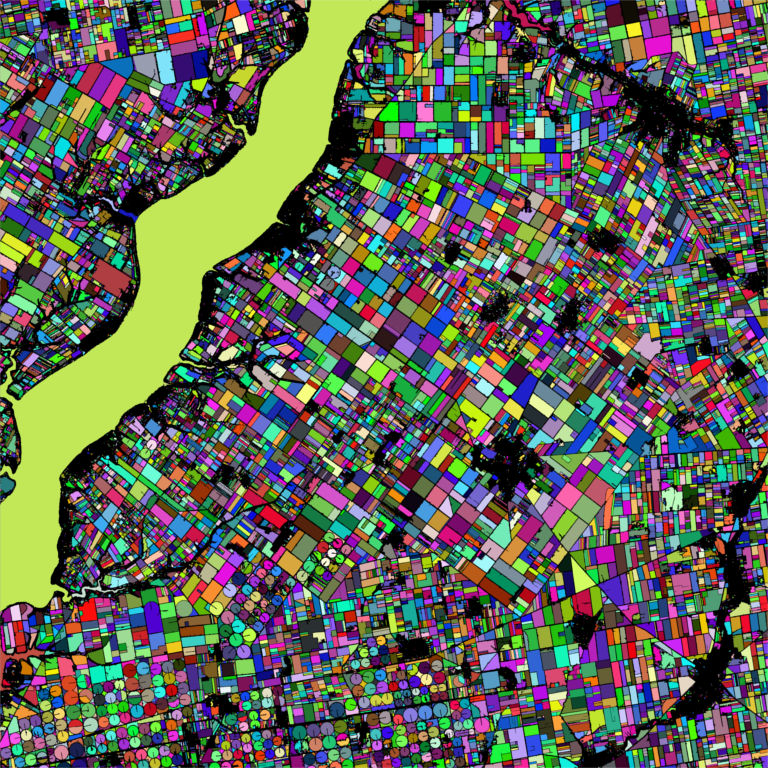}}
  \caption{A segmentation of a portion of southern Ukraine into fields, labeled with random colors. The fieldmapping process makes use of many images from NASA's Landsat 7 and 8 satellites and ESA's Sentinel 2A satellite. Having pixels grouped into fields allows us to exploit strong spatial correlations in our agricultural analysis.}\label{fig:fieldmap}
\end{figure}

\begin{figure*}[bt]
  \centering
  \resizebox{7.08in}{!}{\includegraphics{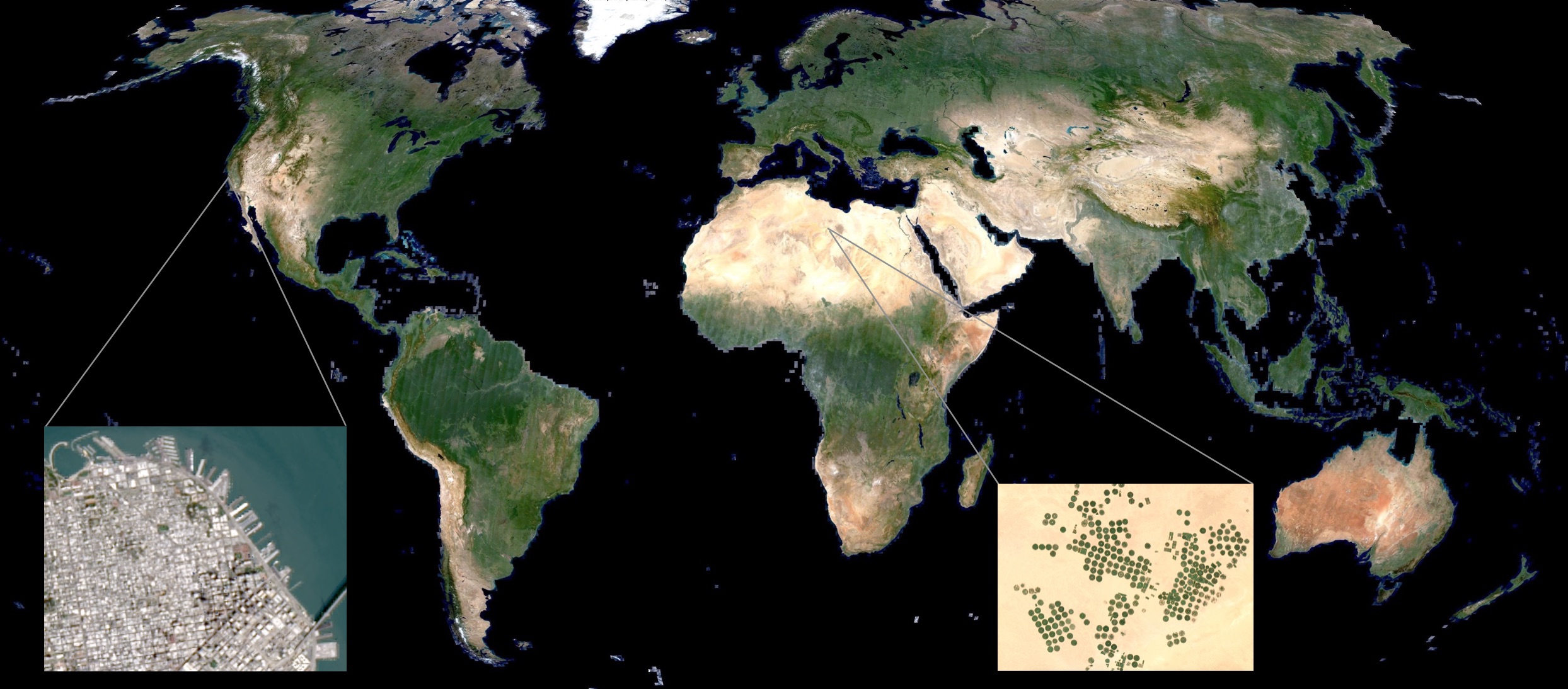}}
  \caption{A cloud-free, 15-meter composite image of the world using all pre-September 2016 Landsat 8 imagery.  The insets show downtown San Francisco and irrigated agriculture in Libya.}
\label{fig:composite}
\end{figure*}

\subsection{Field segmentation}

Analysis of satellite imagery that is done at the pixel level fails to take advantage of an obvious source of additional information. The results of such analysis should be highly spatially correlated, as nearby pixels are likely to have the same land use. In the case of agriculture in particular, the land is divided into fields in which the same crop is planted, and which is subsequently subjected to very similar conditions. Identifying the pixels that belong to the same fields thus leads to improved analysis results.

The image in Figure~\ref{fig:fieldmap} shows the segmentation into fields of one of our UTM tiles, covering a portion of Kherson Oblast in southern Ukraine, bordering the Dnieper River. The tile is 6144 x 6144 pixels at 10 m resolution. The segmentation is produced from multispectral imagery from NASA's Landsat 7 and Landsat 8 satellites collected since 2011, as well as from ESA's Sentinel 2A satellite collected since the beginning of its mission in June 2015. Using a thin API layer on top of Festivus makes it straightforward to process imagery in parallel from multiple different sensors with their own unique tilings in a consistent, uniform way.

The field segmentation process begins with identifying edges. While clouds and other artifacts in any given image can obscure edges and introduce spurious ones, the edges we care about have the property of being persistent in time. Examining temporal edge statistics allows us to use our deep temporal image stack to identify edges of interest. First, for each image we apply a simple cloud mask \cite{oreopoulos-2011-implementation}, and remove cloud pixels from the valid data region. We then compute the spatial gradient magnitude, ensuring that only changes across valid pixels produce nonzero gradients. (For example, this keeps the Landsat 7 scan-line corrector artifacts from producing spurious edges.) The magnitude is accumulated over the bands of each image and over the images available in the chosen time interval, along with a count of how many times each pixel contained valid data. These quantities are divided pixelwise to produce a temporal-mean gradient image, which is then thresholded to produce a binary edge map. Morphological operations are used to clean up the edges, and then the non-edge pixels are separated into connected components. These components are labeled and polygonized, and the resulting polygons stored as a GeoJSON file. For display purposes, assigning a random color to each labeled region allows the individual fields to be visually distinguished.

\subsection{Cloud-free composite images}

We have used the virtual file system described above to create a 15-meter, cloud-free composite image of the world using all pre-September 2016 Landsat 8 data (Figure~\ref{fig:composite}).  The input for this computation consists of 68 TB of JPEG 2000-compressed imagery and spans 3.4 years of earth observation.  The output is a weighted average of this imagery, with higher weight given to cloud-free, verdant input images.

The work was easily parallelized by dividing the earth's surface into 43k square tiles; each tile was processed independently.  The computation was distributed across 400 32-vCPU pre-emptible instances and lasted 8 hours, for a total of 100k CPU-hours and a cost of \$1000.

A similar composite image was announced by the Google Maps team in 2016 \url{https://maps.googleblog.com/2016/06/keeping-earth-up-to-date-and-looking.html}.  We estimate that our computation requires at least 75X less CPU-hours per day of Landsat imagery
than the Google Maps computation.

\section{Conclusion}

While the race to exa-scale explores accelerated architectures with
their associated breakage of established software interfaces, the
cloud offers an alternative convergent architecture which can take
full advantage of economies of scale and continue to protect
application investment over multiple generations of hardware.
While it took over 40 years for Landsat to collect a petabyte of data,
next year Planet's satellite constellation~\cite{boshuizen_results_2014} will
produce many petabytes.  In this work we
have demonstrated that cloud computing architectures are suitable for
at least one important class of data-intensive scientific analysis.

\section{Acknowledgments}

We thank Solomon Boulos, Scott Van Woudenberg and Paul Nash from
Google for guidance and assistance, and the Google Earth Engine
project for providing open access to Landsat data.  We thank Jed Sundwall of
Amazon Web services for facilitating access to Copernicus
Sentinel-2 data.  This publication is supported in part by the Gordon
and Betty Moore Foundation's Data-Driven Discovery Initiative through
Grant GBMF4561 to Matthew Turk.  We thank NASA and USGS for making the
full archives of Landsat and MODIS imagery publicly available, and the
EU Copernicus program for making Sentinel data available.

\bibliographystyle{myabbrvnat}
\bibliography{ms}

\begin{thebibliography}{23}
\providecommand{\natexlab}[1]{#1}
\providecommand{\url}[1]{\texttt{#1}}
\expandafter\ifx\csname urlstyle\endcsname\relax
  \providecommand{\doi}[1]{doi: #1}\else
  \providecommand{\doi}{doi: \begingroup \urlstyle{rm}\Url}\fi

\bibitem[Anderson et~al.(1995)Anderson, Culler, and
  Patterson]{anderson1995case}
T.~E. Anderson, D.~E. Culler, and D.~Patterson.
\newblock A case for {NOW} (networks of workstations).
\newblock \emph{IEEE micro}, 15\penalty0 (1):\penalty0 54--64, 1995.

\bibitem[Bailey et~al.(1991)Bailey, Barszcz, Barton, Browning, Carter, Dagum,
  Fatoohi, Frederickson, Lasinski, Schreiber, Simon, Venkatakrishnan, and
  Weeratunga]{bailey_nas_1991}
D.~H. Bailey et~al.
\newblock The {NAS} {Parallel} {Benchmarks}.
\newblock \emph{International Journal of High Performance Computing
  Applications}, 5\penalty0 (3):\penalty0 63--73, Sept. 1991.
\newblock URL \url{http://hpc.sagepub.com/content/5/3/63}.

\bibitem[Becker et~al.(1995)Becker, Sterling, Savarese, Dorband, Ranawak, and
  Packer]{becker_beowulf:_1995}
D.~J. Becker, T.~Sterling, D.~Savarese, J.~E. Dorband, U.~A. Ranawak, and C.~V.
  Packer.
\newblock {BEOWULF}: {A} parallel workstation for scientific computation.
\newblock In \emph{Proceedings, {International} {Conference} on {Parallel}
  {Processing}}, volume~95, 1995.
\newblock URL
  \url{http://www.phy.duke.edu/~rgb/brahma/Resources/beowulf/papers/ICPP95/icpp95.html}.

\bibitem[Bell and Gray(2002)]{bell_whats_2002}
G.~Bell and J.~Gray.
\newblock What's {Next} in {High}-performance {Computing}?
\newblock \emph{Commun. ACM}, 45\penalty0 (2):\penalty0 91--95, Feb. 2002.
\newblock URL \url{http://doi.acm.org/10.1145/503124.503129}.

\bibitem[Boshuizen et~al.(2014)Boshuizen, Mason, Klupar, and
  Spanhake]{boshuizen_results_2014}
C.~Boshuizen, J.~Mason, P.~Klupar, and S.~Spanhake.
\newblock Results from the {Planet} {Labs} {Flock} {Constellation}.
\newblock \emph{AIAA/USU Conference on Small Satellites}, Aug. 2014.
\newblock URL \url{http://digitalcommons.usu.edu/smallsat/2014/PrivEnd/1}.

\bibitem[Carlson(2013)]{carlson_redis_2013}
J.~L. Carlson.
\newblock \emph{Redis in {Action}}.
\newblock Manning Publications Co., Greenwich, CT, USA, 2013.
\newblock ISBN 1617290858, 9781617290855.

\bibitem[{Celery Development Team}(2015)]{celery}
{Celery Development Team}.
\newblock Celery - the distributed task queue, 2015.
\newblock URL \url{http://www.celeryproject.org}.

\bibitem[ISO/IEC JTC 1/SC 29/WG 1()]{jtc1sc29-2004-jpeg2000}
ISO/IEC JTC 1/SC 29/WG 1.
\newblock {JPEG} 2000 image coding system {ISO/IEC} 15444-1:2004 $|$ {ITU-T}
  {R}ec. {T}.800.
\newblock International Standard, 2004.
\newblock URL \url{http://www.iso.org/iso/catalogue_detail.htm?csnumber=37674}.

\bibitem[McCalpin(1995)]{mccalpin_memory_1995}
J.~D. McCalpin.
\newblock Memory bandwidth and machine balance in current high performance
  computers.
\newblock 1995.
\newblock URL
  \url{http://www.researchgate.net/publication/213876927_Memory_Bandwidth_and_Machine_Balance_in_Current_High_Performance_Computers}.

\bibitem[Moore(1965)]{moore_1965}
G.~E. Moore.
\newblock Cramming more components onto integrated circuits.
\newblock \emph{Electronics}, 38\penalty0 (8):\penalty0 114--117, Apr. 1965.

\bibitem[Nuvolari(2005)]{FM1284}
A.~Nuvolari.
\newblock Open source software development: Some historical perspectives.
\newblock \emph{First Monday}, 10\penalty0 (10), 2005.
\newblock URL \url{http://firstmonday.org/ojs/index.php/fm/article/view/1284}.

\bibitem[Oreopoulos et~al.(2011)Oreopoulos, Wilson, and
  V\'arnai]{oreopoulos-2011-implementation}
L.~Oreopoulos, M.~J. Wilson, and T.~V\'arnai.
\newblock Implementation on {L}andsat data of a simple cloud-mask algorithm
  developed for {MODIS} land bands.
\newblock \emph{IEEE Geoscience and Remote Sensing Letters}, 8\penalty0
  (4):\penalty0 597--601, July 2011.

\bibitem[Sanfilippo and Noordhuis(2010)]{sanfilippo2010redis}
S.~Sanfilippo and P.~Noordhuis.
\newblock Redis, 2010.
\newblock URL \url{http://redis.io}.

\bibitem[Skillman et~al.(2014)Skillman, Warren, Turk, Wechsler, Holz, and
  Sutter]{skillman_dark_2014}
S.~W. Skillman, M.~S. Warren, M.~J. Turk, R.~H. Wechsler, D.~E. Holz, and P.~M.
  Sutter.
\newblock Dark {Sky} {Simulations}: {Early} {Data} {Release}.
\newblock \emph{arXiv:1407.2600 [astro-ph]}, July 2014.
\newblock URL \url{http://arxiv.org/abs/1407.2600}.
\newblock arXiv: 1407.2600.

\bibitem[Taubman and Marcellin(2001)]{taubman-2001-jpeg2000}
D.~S. Taubman and M.~W. Marcellin.
\newblock \emph{{JPEG}2000: Image Compression Fundamentals, Standards and
  Practice}.
\newblock Kluwer Academic Publishers, Norwell, MA, USA, 2001.
\newblock ISBN 079237519X.

\bibitem[Vatsavai et~al.(2006)Vatsavai, Shekhar, Burk, and
  Lime]{vatsavai2006umn}
R.~R. Vatsavai, S.~Shekhar, T.~E. Burk, and S.~Lime.
\newblock {UMN-MapServer}: A high-performance, interoperable, and open source
  web mapping and geo-spatial analysis system.
\newblock In \emph{International Conference on Geographic Information Science},
  pages 400--417. Springer, 2006.

\bibitem[Warren and Wofford(2007)]{warren2007astronomical}
M.~S. Warren and J.~Wofford.
\newblock Astronomical data analysis with commodity components.
\newblock In \emph{Proceedings of the ACM/IEEE Conference on Supercomputing},
  2007.
\newblock Winner of the SC '07 Storage Challenge Award.

\bibitem[Warren et~al.(1997{\natexlab{a}})Warren, Becker, Goda, Salmon, and
  Sterling]{warren97b}
M.~S. Warren, D.~J. Becker, M.~P. Goda, J.~K. Salmon, and T.~Sterling.
\newblock Parallel supercomputing with commodity components.
\newblock In \emph{Proceedings of the International Conference on Parallel and
  Distributed Processing Techniques and Applications
  {(PDPTA{\textquoteright}97)}}, page 1372{\textendash}1381,
  1997{\natexlab{a}}.
\newblock URL
  \url{http://citeseerx.ist.psu.edu/viewdoc/download?doi=10.1.1.51.2019&rep=rep1&type=pdf}.

\bibitem[Warren et~al.(1997{\natexlab{b}})Warren, Salmon, Becker, Goda,
  Sterling, and Winckelmans]{warren_pentium_1997}
M.~S. Warren, J.~K. Salmon, D.~J. Becker, M.~P. Goda, T.~Sterling, and
  W.~Winckelmans.
\newblock Pentium {Pro} {Inside}: {I}. {A} {Treecode} at 430 {Gigaflops} on
  {ASCI} {Red}, {II}. {Price}/{Performance} of \$50/{Mflop} on {Loki} and
  {Hyglac}.
\newblock In \emph{Supercomputing, {ACM}/{IEEE} 1997 {Conference}}, pages
  61--61, Nov. 1997{\natexlab{b}}.

\bibitem[Warren et~al.(1998)Warren, Germann, Lomdahl, Beazley, and
  Salmon]{warren98a}
M.~S. Warren, T.~C. Germann, P.~S. Lomdahl, D.~M. Beazley, and J.~K. Salmon.
\newblock Avalon: an {Alpha/Linux} cluster achieves 10 gflops for \$150k.
\newblock In \emph{Proceedings of the 1998 {ACM/IEEE} conference on
  Supercomputing {(CDROM)}}, Supercomputing '98, page 1{\textendash}11,
  Washington, {DC}, {USA}, 1998. {IEEE} Computer Society.
\newblock ISBN {0-89791-984-X}.
\newblock URL \url{http://dl.acm.org/citation.cfm?id=509058.509130}.

\bibitem[Warren et~al.(2003)Warren, Fryer, and Goda]{warren_space_2003}
M.~S. Warren, C.~L. Fryer, and M.~P. Goda.
\newblock The {Space} {Simulator}: {Modeling} the {Universe} from {Supernovae}
  to {Cosmology}.
\newblock In \emph{Proceedings of the 2003 {ACM}/{IEEE} {Conference} on
  {Supercomputing}}, {SC} '03, pages 30--, New York, NY, USA, 2003. ACM.
\newblock ISBN 1-58113-695-1.
\newblock URL \url{http://doi.acm.org/10.1145/1048935.1050181}.

\bibitem[Warren et~al.(2015)Warren, Brumby, Skillman, Kelton, Wohlberg, Mathis,
  Chartrand, Keisler, and Johnson]{warren2015seeing}
M.~S. Warren et~al.
\newblock Seeing the earth in the cloud: Processing one petabyte of satellite
  imagery in one day.
\newblock In \emph{2015 IEEE Applied Imagery Pattern Recognition Workshop
  (AIPR)}, pages 1--12. IEEE, 2015.

\bibitem[White(2012)]{white2012hadoop}
T.~White.
\newblock \emph{Hadoop: The definitive guide}.
\newblock " O'Reilly Media, Inc.", 2012.

\end{thebibliography}
\end{document}